\documentclass[pdflatex,sn-basic]{sn-jnl}
\pdfoutput=1

\usepackage{graphicx}
\usepackage{amsmath,amssymb,amsfonts}
\usepackage{amsthm}
\usepackage{mathrsfs}
\usepackage{xcolor}
\usepackage{textcomp}
\usepackage{manyfoot}
\usepackage{booktabs}
\usepackage{multirow}
\usepackage{bm}
\usepackage{setspace}
\usepackage{comment}
\usepackage{float}
\usepackage{newunicodechar}
\newunicodechar{−}{-}
\usepackage{algorithm}
\usepackage{algorithmicx}
\usepackage{algpseudocode}

\usepackage{listings}

\begin{document}

\title[Constraining Dark Energy Dynamics ]{Constraining Dark Energy Dynamics in Curved Spacetime with Current Observations}

\author[1]{\fnm{D. Revanth} \sur{Kumar}}\email{2406c9m003@sru.edu.in}

\author*[1]{\fnm{Santosh Kumar} \sur{Yadav}}\email{sky91bbaulko@gmail.com}

\affil[1]{\orgdiv{Department of Mathematics}, \orgname{SR University}, \city{Warangal}, \postcode{506371}, \state{Telangana}, \country{India}}

\abstract{We investigate a dark energy (DE) equation of state (EoS) parametrization in a curved spacetime using current observations. We constrain the model parameters by using observational Hubble data from Cosmic Chronometer (CC), Pantheon Plus SH0ES (PPS), and DESI BAO DR2, along with their reconstructed datasets using an Artificial Neural Network
(ANN). The parameter $\alpha$ is constrained as $\alpha \approx 0.35 (\approx 0.56)$ from original (reconstructed) data. This means reconstruction pushes the model toward a significant deviation from the standard $\Lambda$CDM framework.  
We find that the curvature parameter $\Omega_{k0} = 0.068 \pm 0.029$ at 68\% CL with original data, suggests a slightly open universe, whereas with the reconstruction method, $\Omega_{k0} = -0.131 \pm 0.032$ at 68\% CL suggests a closed universe. This shift in the mean value indicates that the reconstruction method is highly sensitive to curvature. We perform statistical model comparison criteria, namely, AIC and BIC to assess the reliability of our framework.}


\maketitle

\section{Introduction}
\label{sec:intro}\onehalfspacing

The Lambda Cold Dark Matter ($\Lambda$CDM) model has become the standard framework in cosmology following the major discovery of the Universe’s accelerated expansion, as revealed by observations of distant Type Ia supernovae~\cite{riess1998observational, perlmutter1999measurements}. The model extends the general relativity framework, with a cosmological constant ($\Lambda$) as the primary component of dark energy (DE) responsible for the Universe's acceleration. This model treats DE component as a cosmological constant with an equation of state (EoS) parameter, $w=-1$, and CDM as a cold, pressureless dark matter fluid with EoS $w=0$~\cite{carroll2001cosmological}. The model is broadly accepted as it fits well with most observational evidence at different epochs and scales, such as the cosmic microwave background (CMB) anisotropies~\cite{page2003first}, structure formation on large scales, and baryon acoustic oscillations (BAO). Despite its great fit with observations, it faces several challenges with both theoretical and observational aspects. Theoretically, the most well-known challenges are the coincidence problem~\cite{velten2014aspects}, the fine-tuning problem~\cite{burgess2015cosmological}, and small-scale problems. Observationally, some key parameters show a discrepancy when measured from early universe measurements based on this model and other late universe direct measurements.
One of the most prominent and widely discussed discrepancies (called tension) is in the measurement of the Hubble constant $H_0$ when measured by the Planck~\cite{aghanim2020planck} based on $\Lambda$CDM and local SH0ES collaboration~\cite{riess2022comprehensive} by the distance ladder approach. These issues motivate researchers to investigate alternative cosmological models~\cite{di2021realm} in the light of recent observations.

Further, the Universe is assumed to be homogeneous and isotropic at large scales, but this does not necessarily imply it is flat. Even though most of the observational constraints are consistent with a flat Universe, they still allow a few percent difference from the flatness~\cite{di2021snowmass2021}. The Planck 2018 power spectrum analyses based on the baseline \textit{Plik} likelihood~\cite{handley2021curvature,di2020planck,aghanim2020planck,di2021investigating} indicate that the Universe prefers to closed geometry with more than 3$\sigma$. Additionally, the alternative \textit{CamSpec} likelihood~\cite{efstathiou2020evidence,
efstathiou2019detailed} prefers to closed Universe with more than a 99\% CL. In addition,~\cite{Yang:2022kho} examined extended cosmological models with a free neutrino sector and varying DE parametrizations using Planck 2018, BAO, and Pantheon data, and also found indications of a closed Universe, suggesting that assuming flatness may bias DE constraints. The authors in~\cite{glanville2022full} tested the full-shape theory of curvature using the effective field theory of large-scale structure and found the possibility for non-zero curvature. The authors in~\cite{liu2022revising, yadav2024testing} investigated several cosmological models with different datasets and indicate a slight preference for a non-flat Universe. Using non-CMB datasets~\cite{Wu:2024faw} found a mild preference for an open Universe in the $\Lambda$CDM model, indicating that non-flat models may provide a better fit to current observations. Ref.~\cite{Dias:2024tpf} examined the flatness assumption of the Universe using non-parametric null tests using low-redshift data and found no significant deviation from the flat $\Lambda$CDM scenario. An analysis of $\Lambda$CDM and $X$CDM models show mild evidence for open geometry and quintessence-like behavior, while the flat $\Lambda$CDM model remains most consistent with observations~\cite{deCruzPerez:2024shj}. The author in~\cite{Patil:2022ejk} investigated the interaction in DM-DE in a curved FLRW space–time using a dynamical systems approach and showed that curvature alters cosmological evolution and stability behavior. In a related observational analysis~\cite{Patil:2024mno}, the author investigated spatial curvature in an interacting dark sector model using multiple datasets and found that it influences $H_0$ and structure-growth constraints while indicating a mild preference for an open Universe. Furthermore, the author in~\cite{Yadav:2025wbc} examined the $w$CDM model with spatial curvature using low redshift probes, reporting slight deviations from the cosmological constant and a mild preference for an open geometry. These results suggest that curvature and DE dynamics can be partially degenerate, motivating dedicated studies that allow both to vary simultaneously.
This also indicates that the possibility of a non-flat Universe can not be ruled out, and therefore, it is necessary to investigate the models in which curvature is not fixed to zero.

In recent years, reconstruction techniques have emerged as powerful tools for inferring the cosmic expansion history directly from observations in a model-independent manner~\cite{Chen:2024rzg,Dialektopoulos:2021wde}. These approaches help reduce the impact of observational noise and improve the reliability of cosmological constraints, especially in sparsely sampled redshift regions~\cite{Chen:2022oxg,Wang:2020sxl,Zhang:2023ucf,Wang:2019vxv}. 
Reconstruction techniques have emerged as complementary tools for inferring the cosmic expansion history directly from observations in a model-independent manner. In this context, the ANN-based reconstruction should be viewed as augmenting rather than replacing standard likelihood-based parameter estimation. It provides an independent validation of inferred cosmological trends by testing their stability under controlled smoothing of the expansion history, strengthening confidence that the resulting physical conclusions are not artifacts of measurement scatter.

To achieve this, we derive observational constraints on the DE EoS parametrization proposed in Ref.~\cite{singh2024new}, characterized by a single parameter $\alpha$, within a curved spacetime framework. The analysis uses low-redshift probes including 31 Hubble parameter measurements~\cite{simon2005constraints, stern2010cosmic, Moresco:2012jh, moresco2015raising, moresco20166, zhang2016test, Ratsimbazafy:2017vga}, 1701 Pantheon+SH0ES supernova data points~\cite{brout2022pantheon+}, and 9 DESI BAO measurements from Data Release 2~\cite{karim2025desi}. To ensure the robustness of the results, we perform our analysis using both the original and reconstructed datasets, allowing us to verify that the inferred cosmological evolution is not biased by measurement fluctuations.   The primary objective of this work is to examine whether a smoothly evolving DE component, analyzed within a non-flat spacetime framework, produces observable signatures distinct from those predicted by the standard flat $\Lambda$CDM model. Our analysis evaluates the sensitivity of spatial curvature constraints to the adopted EoS dynamics and determines whether reconstructing observational data modifies parameter inference or reveals hidden degeneracies and systematic trends between curvature and DE evolution.

The following is the structure of the paper: The basic cosmological equations of derived model are presented in Section~\ref {section2}. The observational datasets and methodology parameter estimation is described in Section~\ref {section3}. The Section~\ref{section4} discusses results, while Section~\ref{section5} concludes the paper by summarizing the main findings on the analysis.

\section{Theoretical Model and Basic Equations}\label{section2}
\label{the_model}
Following the cosmological principle, we consider a homogeneous and isotropic Universe modeled by the Friedmann-Lema\^{i}tre-Robertson-Walker (FLRW) with curved spacetime given by the following metric ( without loss of generality, we adopt speed of light $c=1$)
\begin{equation}
    ds^2 = -dt^2 + a^2(t)\bigg[\frac{dr^2}{1-kr^2} + r^2 (d\theta^2 + \sin^2 \theta d\phi^2)\bigg],
\end{equation}
where $a(t)$ denotes the Universe scale factor. The open, flat, and closed geometries of the Universe correspond to spatial curvature k as -1, 0, and +1, respectively. Taking into account the aforementioned metric, the fundamental governing equations, also known as the Friedmann equations, that establish the universe's background evolution, are provided by
\begin{align}\label{model}
H^2 &=  \frac{8 \pi G}{3} (\rho_{\rm i})-\frac{k}{a^2},\\ 
2 \frac{dH}{dt} + 3 H^2 &= - 8 \pi G (p_{\rm i})-\frac{k}{a^2},
\end{align}
where the Hubble parameter $H = \frac{da/dt}{a}$  represents the Universe expansion rate and G represents the Newtonian gravitational constant. 

The pressure and density of $i^{th}$ species are indicated by $p_i$ and $\rho_i$ in the aforementioned formulas, where $ i \in \{ \rm r, \rm m, \Lambda, \rm k\}$ represents radiation, dark matter, cosmological constant, and curvature accordingly.
Under the assumption of no mutual interaction between the above-mentioned species except the usual gravitational interaction, the fluid equation reads as
\begin{align} \label{continiuty}
    \dot{\rho_i} = -  3H\rho_i (1+w_i),
\end{align}
where $w_i$ is the EoS of the $i ^{th}$ specie with the relation $p_i =w_i \rho_i$. As eqn.~(\ref{continiuty}) is satisfied by each component of the Universe, therefore by using eqn.~(\ref{continiuty}), one can write energy density evolution of any $i^{th}$ specie with known EoS, $w_i$.
Using the scale factor, the  new DE EoS parametrization with parameter $\alpha$ is defined as:
\begin{align}\label{EoS}
\omega(a) = -1 + \frac{a^{-\alpha} e^{-2a\alpha} \alpha \, \left(\arctan(a)\right)^{-\alpha}}{3\left(1 + a^{-2\alpha}\right)}.
\end{align}

One may easily write the aforementioned parametrization of DE EoS in terms of redshift using the relation $a=a_0/(1+z)$. Taking $a_0 =1$ at present, the eqn.~(\ref{EoS}) is modified as
\begin{align}\label{EoS interms of Z}
\omega(z) = -1 + \frac{(1 + z)^\alpha e^{\frac{-2\alpha}{(1 + z)}} \alpha \, \left(\arctan(1 + z)\right)^\alpha}{3\left(1 + (1 + z)^{2\alpha}\right)},
\end{align}

Here, the parameter $\alpha$ quantifies the deviation from the cosmological constant behavior.  The parameter $\alpha$ plays a crucial role in governing the redshift evolution of the dark
energy equation of state in the proposed parametrization. Although it has been introduced as a phenomenological parameter, its sign and magnitude directly control the deviation of the model from the standard cosmology framework and therefore possess clear theoretical implications.
Positive values of $\alpha$ correspond to a dynamically evolving DE component. At $\alpha=0$, this parameterization reproduces the $\Lambda$CDM model. However, for arbitrary values of $\alpha$, at large redshifts ($z \to \infty$), the equation of state parameter approaches to $ -1$. This indicates that the model effectively converges to the $\Lambda$CDM behavior in the early Universe, thereby ensuring that the Big Bang nucleosynthesis bounds are automatically satisfied.

\begin{align}\label{DE Evolution}
\frac{\rho_{de}}{\rho_{de0}} = e^{\frac{\alpha z}{1+z}} \, \frac{\left(\arctan(1 + z)\right)^\alpha}{\arctan 1},
\end{align}
where $\rho_{de0}$ is the present DE density. Now, the Friedmann equation can be reformulated using density parameters as 
\begin{align}\label{Novel Model}
E^2 (z) = \frac{H(z)}{H_0}=  \Omega_{m0} (1 + z)^3 +\Omega_{k0} (1+z)^2 + \Omega_{de0} e^{\frac{\alpha z}{1+z}} \, \frac{\left(\arctan(1 + z)\right)^\alpha}{\arctan 1},
\end{align}
where the Hubble parameter's current value is denoted by $H_0$. The quantities: $\mathrm{\Omega}_{m0},~ \mathrm{\Omega}_{de0},~ \mathrm{\Omega}_{k0}$ are density parameters of matter, DE, and curvature, respectively, following the total budgets equation, $\mathrm{\Omega}_{m0}+\mathrm{\Omega}_{de0}+\mathrm{\Omega}_{ko0}=1$. The first term, ${(1\ +\ z)}^3$ reflect the matter content evolution, second term ${(1\ +\ z)}^2$ shows for the curvature evolution, and the final term $\rho_{de}/\rho_{de0}$ indicate the DE evolution for considered parametrization.

\section{Observational Data Analysis}\label{section3}

In this analysis, multiple observational datasets are utilized to constrain the model parameters. We have adopted two different approaches for constraining the model parameters. Firstly, we have applied the Bayesian MCMC method with original (actual) data from different measurements, and then we have performed Artificial Neural Network (ANN)-based reconstruction of the same sets of data. The constraints are placed on model parameters from both the original and reconstructed data.
We discuss below the details of the original observational datasets, followed by the ANN-based reconstruction of observational data.
\subsection{Data Description}
\subsubsection{Cosmic Chronometers (CC)} 
The CC approach is an efficient method for tracing the cosmic expansion history of the universe, and measures the Hubble parameter $H(z)$ at various redshifts.
We employ 31 CC data points in the redshift range from $z=0.07$ to $z=1.965$, from multiple measurements~\cite{simon2005constraints, stern2010cosmic, Moresco:2012jh, moresco2015raising, moresco20166, zhang2016test, Ratsimbazafy:2017vga}. The data points are summarized in Table~\ref{OHDdata31}. We refer to this dataset as \texttt{CC} in the forthcoming text. 
\begin{table}[h]
\caption{The 31 $H(z)$ measurements from the CC method used in this study in units of $\mathrm{Km\,s^{-1} Mpc^{-1}}$.}
\label{OHDdata31}
\begin{tabular}{l c c c c}
            \hline\hline
            S.No & $z$ & $H(z)$ & Error  & Reference \\
            \hline
                1  & 0.07    & 69.0  & 19.6 & \cite{zhang2016test} \\
                2  & 0.09    & 69.0  & 12.0 & \cite{simon2005constraints} \\
                3  & 0.12    & 68.6  & 26.2 & \cite{zhang2016test} \\
                4  & 0.17    & 83.0  & 8.0  & \cite{simon2005constraints} \\
                5  & 0.179   & 75.0  & 4.0  & \cite{Moresco:2012jh} \\
                6  & 0.199   & 75.0  & 5.0  & \cite{Moresco:2012jh} \\
                7  & 0.20    & 72.9  & 29.6 & \cite{zhang2016test} \\
                8  & 0.27    & 77.0  & 14.0 & \cite{simon2005constraints} \\
                9  & 0.28    & 88.8  & 36.6 & \cite{zhang2016test} \\
                10 & 0.352   & 83.0  & 14.0 & \cite{Moresco:2012jh} \\
                11 & 0.3802  & 83.0  & 13.5 & \cite{moresco20166} \\
                12 & 0.40    & 95.0  & 17.0 & \cite{simon2005constraints} \\
                13 & 0.4004  & 77.0  & 10.2 & \cite{moresco20166} \\
                14 & 0.4247  & 87.1  & 11.2 & \cite{moresco20166} \\
                15 & 0.4497  & 92.8  & 12.9 & \cite{moresco20166} \\
                16 & 0.47    & 89.0  & 50.0 & \cite{Ratsimbazafy:2017vga} \\
                17 & 0.4783 & 80.9  & 9.0   & \cite{moresco20166} \\
                18 & 0.48   & 97.0  & 62.0  & \cite{stern2010cosmic} \\
                19 & 0.593  & 104.0 & 13.0  & \cite{Moresco:2012jh} \\
                20 & 0.68   & 92.0  & 8.0   & \cite{Moresco:2012jh} \\
                21 & 0.781  & 105.0 & 12.0  & \cite{Moresco:2012jh} \\
                22 & 0.875  & 125.0 & 17.0  & \cite{Moresco:2012jh} \\
                23 & 0.88   & 90.0  & 40.0  & \cite{stern2010cosmic} \\
                24 & 0.9    & 117.0 & 23.0  & \cite{simon2005constraints} \\
                25 & 1.037  & 154.0 & 20.0  & \cite{Moresco:2012jh} \\
                26 & 1.3    & 168.0 & 17.0  & \cite{simon2005constraints} \\
                27 & 1.363  & 160.0 & 33.6  & \cite{moresco2015raising} \\
                28 & 1.43   & 177.0 & 18.0  & \cite{simon2005constraints} \\
                29 & 1.53   & 140.0 & 14.0  & \cite{simon2005constraints} \\
                30 & 1.75   & 202.0 & 40.0  & \cite{simon2005constraints} \\
                31 & 1.965  & 186.5 & 50.4  & \cite{moresco2015raising} \\
\hline\hline
\end{tabular}
\end{table}

The likelihood function for the CC is defined as
\begin{align}\label{chi^2}
\chi^2_{\rm CC}(z) = \sum_{i=1}^{31} \left[\frac{H_{\text{t}}\left(\alpha, H_0, z_i\right) - H_{\text{ob}}\left(z_i\right)}{\sigma_H\left(z_i\right)}\right]^2,
\end{align}
where $H_{\text{t}}\left(\alpha, H_0, z_i\right)$ is the theoretical value determined from our considered model at different redshifts $z_i$, and $H_{\text{ob}}\left(z_i\right)$ corresponds to the observed Hubble parameter, while $\sigma_H$ indicates the measurement error.

\subsubsection{Pantheon Plus and SH0ES} 
Type Ia supernovae (SNe Ia) are widely employed as standard candles owing to their consistent peak luminosity. We utilize the distance moduli data from SNe Ia. 
This dataset includes 1701 light curves corresponding to 1550 different
SNe Ia events distributed across the redshift range from $z=0.001$ to $z=2.26$. The theoretical apparent magnitude ($m_B$) of a  supernova at redshift $z$ is given below
\begin{eqnarray}
\label{distance_modulus}
m_B = 5 \log_{10} \left[ \frac{d_L(z)}{1\rm Mpc} \right] + 25 + M_B,
\end{eqnarray}
Where $M_B$ denotes absolute magnitude and $d_L(z)$ represents luminosity distance.
\begin{equation}
    d_L(z)=\frac{\left(1+z\right)}{H_0}\int_{0}^{z}\frac{dx}{E(x)},
\end{equation}
where $E(x)$ can be obtained from eqn.~(\ref{Novel Model}).

We use SNe Ia distance modulus data from the Pantheon Plus sample~\cite{brout2022pantheon+}. Additionally, the SH0ES Cepheid host distance anchors~\cite{brout2022pantheon+} are used in this study, and we refer to this combined dataset as \texttt{PPS}. The likelihood function is defined as
\begin{eqnarray}
\label{chi^2 for PPS}
\chi^2_{\rm PPS}(z)=\sum_{j,k=1}^{1701}{\mathrm{\nabla}_{\mu_j}\left(C_{SN}^{-1}\right)_{j,k}}\mathrm{\nabla}_{\mu_k},
\end{eqnarray}
where $\mathrm{\nabla}_\mu=\mu_{th}-\mu_{obs}$ represents the residual between the theoretical $\mu_{th}$ and observed $\mu_{obs}$ distance modulus and $C_{SN}$ is the covariance matrix consisting systematic and statistical errors which we can get from~\cite{PPSdata}. Theoretically, we can calculate the distance modulus function using
\begin{equation}
    \mu_{th}=5\,\log{\left(d_L\right)}+25.
\end{equation}

\subsubsection{DESI BAO DR2}
The density of visible baryonic matter exhibits periodic and recurrent variations, referred to as BAO. These oscillations serve as crucial standard rulers for accurately measuring distances in the field of cosmology. The BAO scale can be expressed as the sound horizon at the drag epoch $r_d$, the distance a sound wave would have traveled before baryon-photon decoupling.
The drag epoch occurs after photon decoupling due to the residual interaction of baryons with photons, as baryons are significantly fewer in number compared to photons. The drag scale $r_d$ is determined as:
\begin{equation}
  r_d = \int_{z_d}^{\infty} \frac{c_s(z)}{H(z)} \, dz
\end{equation}
where $c_s(z)$ represents the baryon-photon plasma's sound speed, and $z_d$  represents the drag epoch's redshift.

The DESI is an advanced spectroscopic survey to improve BAO constraints by tracing large-scale matter distribution. DESI employs multiple matter tracers to measure BAO over a range from $0.3$ to $2.33$. The isotropic BAO datasets provide measurements 
from Bright Galaxy Survey (BGS). The anisotropic BAO measurements come from Luminous Red Galaxies (LRGs), Emission Line Galaxies (ELGs), and Quasi-Stellar Objects (QSOs), and Lyman-$\alpha$ quasars (Lya QSOs). 
In the second data release (DR2), the DESI survey has been significantly increased, covering 6671 dark tiles and 5171 bright tiles. This corresponds to an improvement by a factor of about 2.4 for the dark program and 2.3 for the bright program compared to DR1~\cite{adame2024desi}. Using the BAO measurements from DESI DR2, we can constrain the combined parameter $H_0 r_d$, but not $H_0$ or $r_d$ individually.
In this work, we have considered DESI BAO DR2 observations from~\cite{karim2025desi} as mentioned here in Table~\ref{DESI BAO}. We refer to this data simply as \texttt{DESI DR2} in the remaining text.

\begin{table}[ht]
\caption{The 9 points from DESI BAO DR2 measurements used in the present analysis in units of $\mathrm{Km\,s^{-1} Mpc^{-1}}$. }
\label{DESI BAO}
\begin{tabular}{l c c c c c r}
 
				\hline\hline
				{tracer} &  $z_{\rm eff}$  \,\,\,   &  \,\,\,   $D_M/r_d$ \,\,\,  &\,\,\,  $D_H/r_d$ \,\,\, & \,\,\, $D_V/r_d$ \,\,\, 
				\\
				\hline 
				{\scriptsize {BGS }}  &  $0.295$ & $-$ &$-$ & $7.942 \pm 0.075$
				\\
				{\scriptsize {LRG1}}  &  $0.510$ & $13.588 \pm 0.167$ & $ 21.863 \pm 0.425$ & $12.720 \pm 0.099$
				\\
				{\scriptsize {LRG2}} &  $0.706$ & $17.351 \pm 0.177$ & $19.455 \pm 0.330$ & $16.050 \pm 0.110$
				\\
				{\scriptsize {LRG3} } &  $0.922$ & $21.648 \pm 0.178$ & $17.577 \pm 0.213$ & $19.656 \pm 0.105$
                \\
                {\scriptsize {ELG1} } &  $0.955$ & $21.707 \pm 0.335$ & $17.803 \pm 0.297$ & $20.008 \pm 0.183$
                \\
                {\scriptsize {LRG3+ELG1} } &  $0.934$ & $21.576 \pm 0.152$ & $17.641 \pm 0.193$ &$19.721 \pm 0.091$
                \\
                
                {\scriptsize {ELG2} } &  $1.321$ & $27.601 \pm 0.318$ & $14.176 \pm 0.221$ & $24.252 \pm 0.174$
                \\
                {\scriptsize {QSO} } & $1.484$ & $30.512 \pm 0.760$ &$12.817 \pm 0.516$ & $26.055 \pm 0.398$ 
                \\
                {\scriptsize {Ly-$\alpha$ QSO} } &  $2.330$ & $38.988 \pm 0.531$ & $8.632 \pm 0.101$ & $31.267 \pm 0.256$
				\\
				\hline\hline  
			\end{tabular}    
\end{table}
    
\subsubsection{ANN-based Reconstruction of Observational Data}
To reconstruct the observational data, we employ a feed-forward neural network (multilayer perceptron) architecture consisting of two hidden layers, each with 64 neurons activated by the Rectified Linear Unit (ReLU) function~\cite{chen2020dynamic}, and a single output neuron that predicts the observable corresponding to each redshift. We implemented \texttt{TensorFlow 2.x} with the \texttt{tensorflow.keras} API in Python.

To incorporate observational uncertainties, we adopt a \texttt{bootstrap resampling approach}. Several resampled datasets are generated by resampling data points with replacement from the original data. Each bootstrap sample is used to train a separate ANN, using a weighted mean-squared error loss where the weights are inversely proportional to the square of the measurement uncertainties. The Predictions are then made at the original redshift points. This procedure is carried out over many bootstrap realizations to generate a distribution of predicted values at each redshift. The median of these predictions is taken as the reconstructed value, and its uncertainty is obtained from the 16th and 84th percentiles of the bootstrap predictions.

\subsection{Methodology}

In modern cosmology, MCMC methods have become essential for probing and analyzing complex parameter spaces within a Bayesian framework. These techniques provide an efficient method for sampling from probability distributions, enabling the robust estimation of model parameters along with their associated uncertainties. We have performed MCMC analysis utilising both original data and ANN-based reconstructed data for the derived model of the universe. We used the \texttt{emcee} Python package~\cite{foreman2013emcee} to obtain the sample on the model parameters. The \texttt{GetDist} module in Python is used to examine the MCMC samples.
We have used flat priors on the model parameter, which are as follows: $H_0 \in [40,100]$,  $\Omega_{m0} \in [0,1]$, $\Omega_{k0} \in [-1,1]$, and  $\alpha \in [-5,3]$.

\section{Results and Discussion}\label{section4}
\begin{figure}[ht]
{ \includegraphics[scale=0.35]{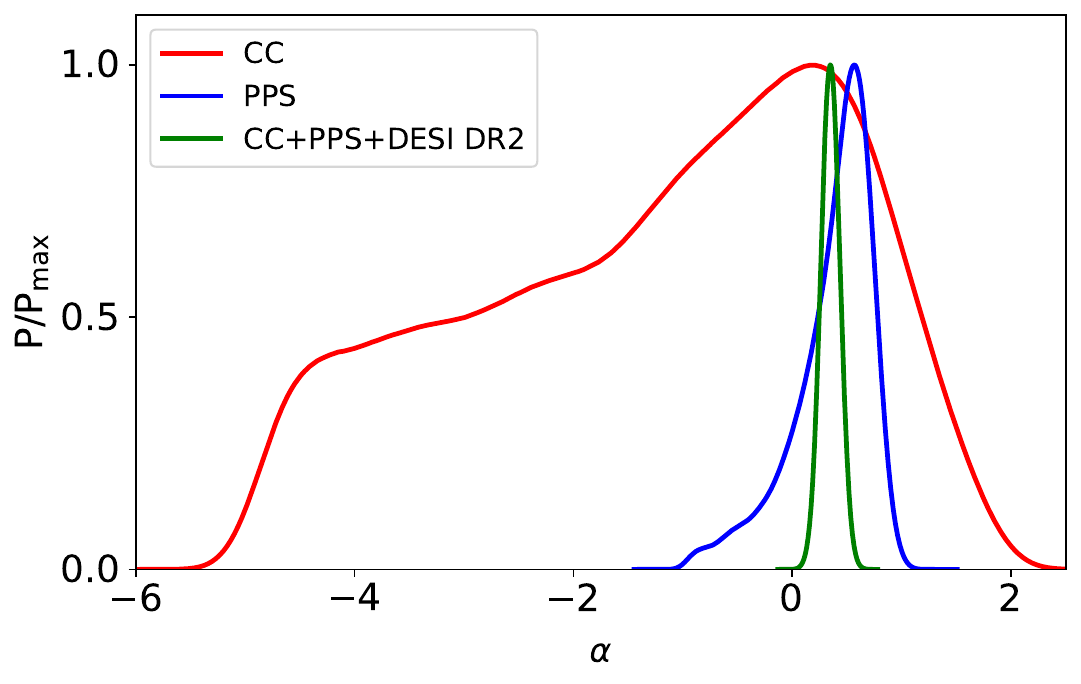}}
	 { \includegraphics[scale=0.35]{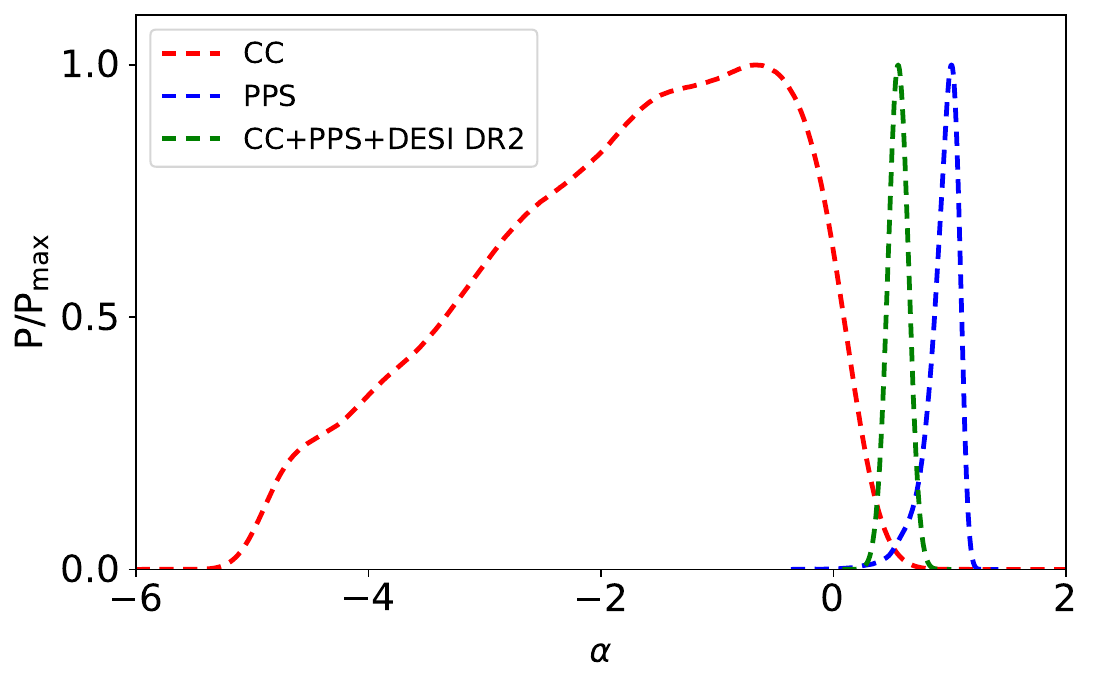}}\\
	 \caption{One-dimensional marginalized probability distribution for parameter $\alpha$ of our model using original (left panel) and reconstructed  (right panel) dataset.}
\label{1-dim_alpha}
\end{figure}

\begin{figure}[ht]
{ \includegraphics[scale=0.35]{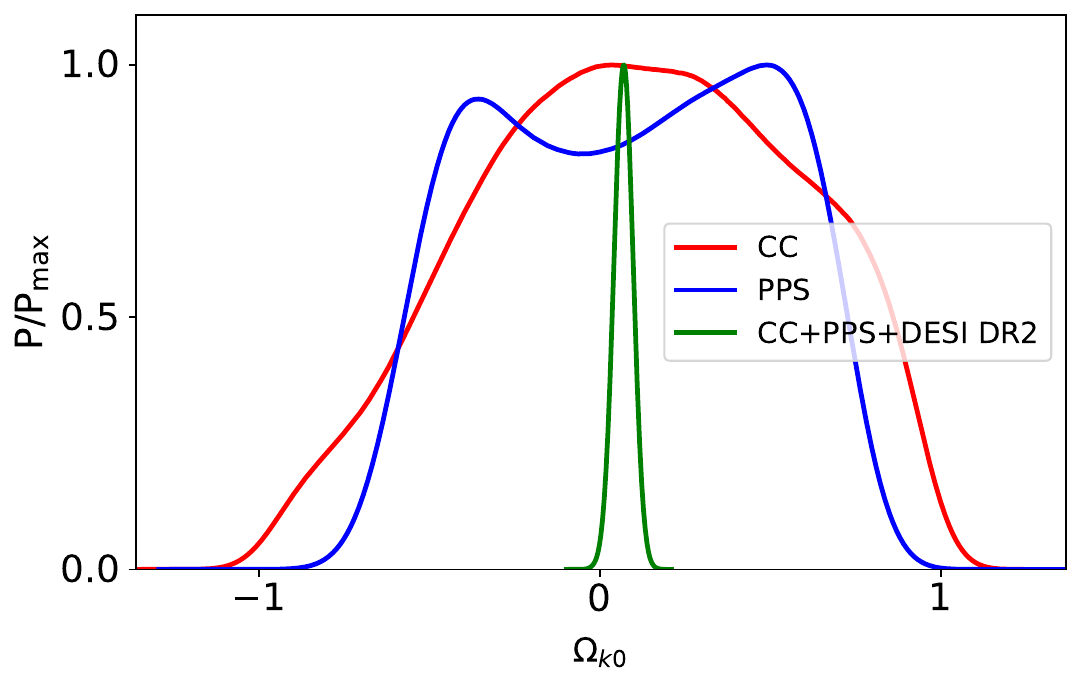}}
	 { \includegraphics[scale=0.35]{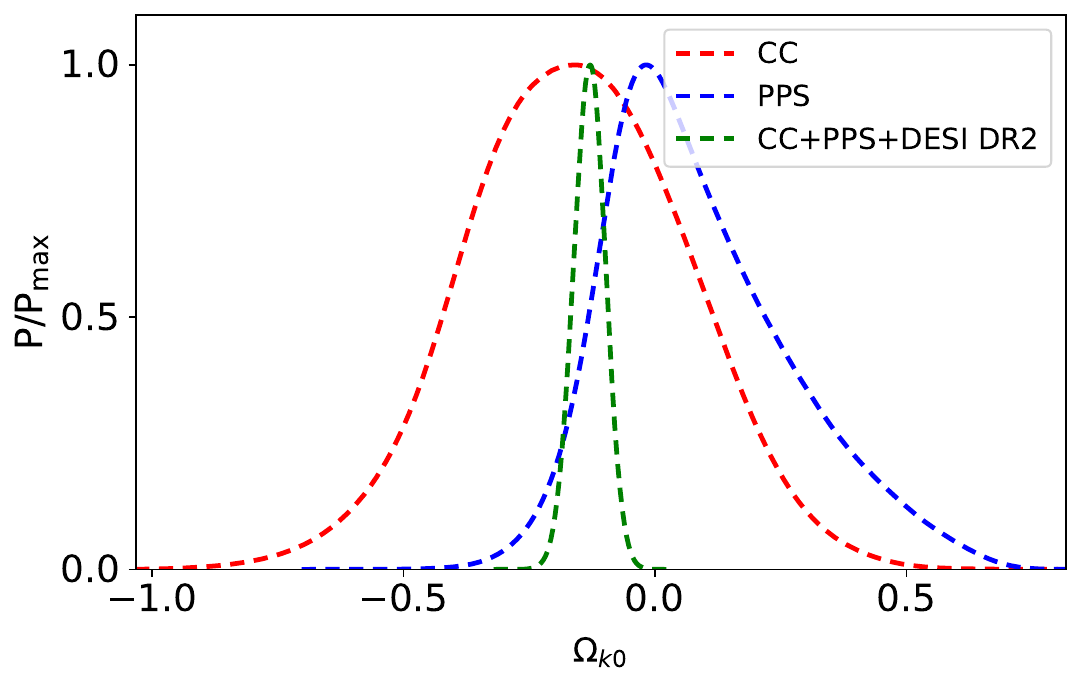}}\\
	 \caption{One-dimensional marginalized probability distribution of $\Omega_{k0}$ for our model using original (left panel) and reconstructed (right panel) dataset.}
\label{1-dim}
\end{figure}

In this section, we present the best-fit values obtained for our model parameters using the following datasets: CC, PPS, and CC + PPS + DESI DR2. 
We have derived constraints on parameters from individual data sets, namely, with CC and PPS, and we observe that the resulting constraints are weaker with large errors. It is evident from the one-dimensional marginalized distributions of $\alpha$ in Figure~\ref{1-dim_alpha} and $\Omega_{k0}$ in Figure~\ref{1-dim}. It is important to mention that there is no degeneracy between these data sets. Therefore, to obtain statistically significant and reliable limits on the cosmological parameters, we combined these two datasets with DESI DR2 and performed the joint analysis: CC+PPS+DESI DR2. The resulting constraints from joint analysis at 68\%, 95\%, and 99\% CL for both original and reconstructed datasets are displayed in Table~\ref{tabparm1}.
We observe that the free parameter $\alpha$ deviates from zero for both the original and reconstructed datasets, indicating a mild departure from the standard $\Lambda$CDM model, see Table~\ref{tabparm1}. For the original dataset, the best-determined limit is $\alpha = 0.348 \pm 0.087$ at 68\% CL, whereas for the reconstructed dataset, the mean value increases to $\alpha = 0.557 \pm 0.089$ at 68\% CL, with joint analysis. These positive values of $\alpha$ consistently suggest the possible dynamical nature of DE in the derived model. One may also see the one-dimensional distributions of $\alpha$ for the original (left panel) and reconstructed (right panel) datasets in Figure~\ref{1-dim_alpha}.

\begin{table}[ht]
\caption{Constraints on the cosmological parameters for our model, $\Lambda$CDM, and $\Lambda$CDM + $\Omega_k$ models using the CC+PPS+DESI DR2 for both original and reconstructed datasets. The Hubble constant is measured in the units of $\mathrm{Km\,s^{-1} Mpc^{-1}}$.}
\label{tabparm1}
\renewcommand{\arraystretch}{1.6}
\begin{tabular}{l l c c c c}
\hline\hline
\textbf{Model} & \textbf{Data} & \textbf{Parameter} & \textbf{68\% limits} & \textbf{95\% limits} & \textbf{99\% limits} \\
\hline

\multirow{8}{*}{Our Model} 
& \multirow{4}{*}{Original} 
& $H_0$         & $68.92 \pm 0.32$                               & $68.92 \pm 0.63$            & $68.92^{+0.84}_{-0.82}$ \\
&               & $\Omega_{m0}$ & $0.277 \pm 0.027$              & $0.277^{+0.054}_{-0.052}$   & $0.277^{+0.071}_{-0.068}$ \\
&               & $\Omega_{k0}$ & $0.068 \pm 0.029$              & $0.068^{+0.055}_{-0.056}$   & $0.068^{+0.072}_{-0.075}$ \\
&               & $\alpha$      & $0.348 \pm 0.087$              & $0.35 \pm 0.17$              & $0.35 \pm 0.22$ \\
\cmidrule{2-6}
& \multirow{4}{*}{Reconstructed} 
& $H_0$         & $68.80 \pm 0.36$               & $68.80^{+0.72}_{-0.71}$     & $68.80 \pm 0.94$ \\
&               & $\Omega_{m0}$ & $0.395 \pm 0.035$              & $0.395^{+0.070}_{-0.069}$   & $0.395^{+0.092}_{-0.090}$ \\
&               & $\Omega_{k0}$ & $-0.131 \pm 0.032$             & $-0.131^{+0.063}_{-0.064}$  & $-0.131^{+0.082}_{-0.084}$ \\
&               & $\alpha$      & $0.557 \pm 0.089$              & $0.56^{+0.17}_{-0.18}$      & $0.56 \pm 0.23$ \\
\hline

\multirow{4}{*}{$\Lambda$CDM} 
& \multirow{2}{*}{Original} 
& $H_0$         &  $73.40\pm 0.15$ &  $73.40 \pm 0.29$     &  $73.40^{+0.38}_{-0.39}$ \\
&               & $\Omega_{m0}$ &  $0.3056\pm 0.0073$            &  $0.306^{+0.015}_{-0.014}$   &  $0.306^{+0.019}_{-0.018}$ \\
\cmidrule{2-6}
& \multirow{2}{*}{Reconstructed} 
& $H_0$         & $72.94 \pm 0.15$               & $72.94 \pm 0.30$     & $72.94 \pm 0.39$ \\
&               & $\Omega_{m0}$ & $0.286 \pm 0.0072$              & $0.286 \pm 0.014$   & $0.286^{+0.019}_{-0.018}$ \\

\hline

\multirow{6}{*}{ $\Lambda$CDM + $\Omega_k$} 
& \multirow{3}{*}{ Original}
&  $H_0$         &  $73.33\pm 0.16$ &  $73.33 \pm 0.31$     &  $73.33^{+0.41}_{-0.40}$ \\
&               &  $\Omega_{m0}$ &  $0.331\pm 0.020$             &  $0.331^{+0.040}_{-0.038}$   &  $0.331^{+0.053}_{-0.050}$ \\
&               &  $\Omega_{k0}$ &  $-0.031\pm 0.023$              &  $-0.031^{+0.045}_{-0.046}$   &  $-0.031^{+0.059}_{-0.060}$\\
\cmidrule{2-6}
& \multirow{3}{*}{ Reconstructed}
&  $H_0$         &  $72.51\pm 0.16$ &  $72.51 \pm 0.32$     &  $72.51 \pm 0.42$ \\
&               &  $\Omega_{m0}$ &  $0.546\pm 0.024$              &  $0.546^{+0.048}_{-0.046}$   &  $0.546^{+0.063}_{-0.060}$ \\
&               &  $\Omega_{k0}$ &  $-0.326\pm 0.026$              &  $-0.326^{+0.050}_{-0.051}$   &  $-0.326^{+0.066}_{-0.067}$ \\

\hline\hline

\end{tabular}
\end{table}

\begin{figure}[ht]
\includegraphics[scale=0.48]{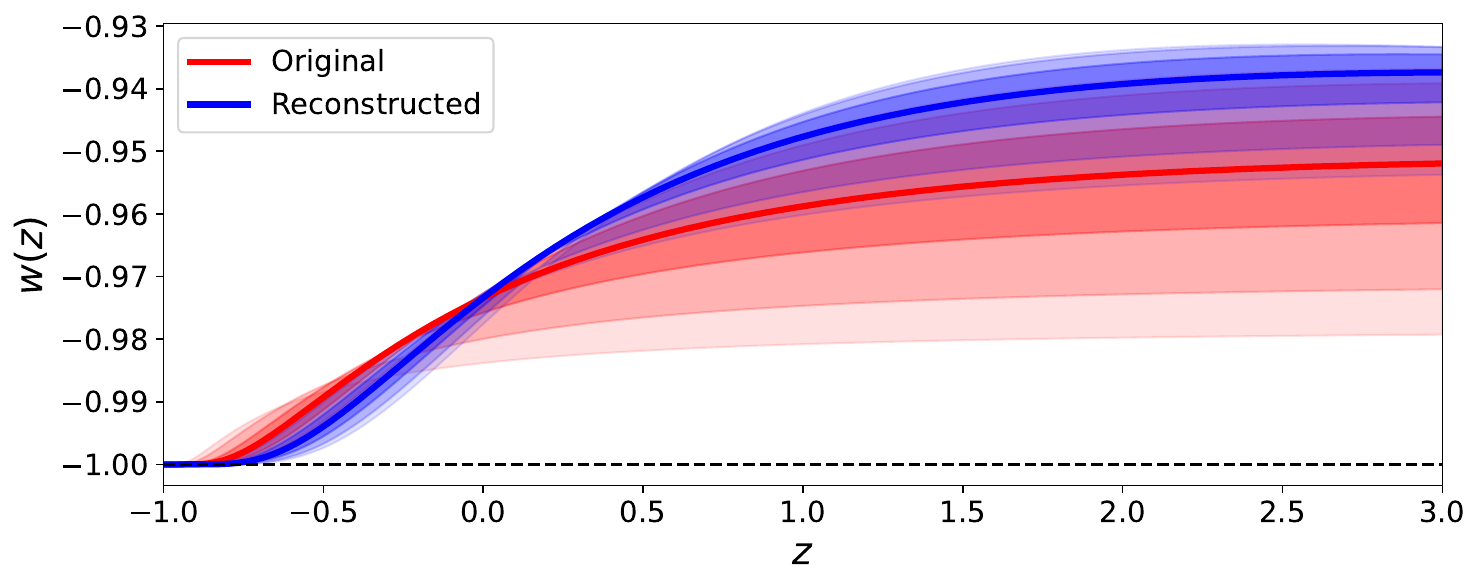}
	 \caption{Variation of the dark energy equation of state parameter $w(z)$ with redshift $z$ of our model using original and reconstructed datasets with joint analysis with  1$\sigma$, 2$\sigma$, and 3$\sigma$ CL.}
\label{w(z)}
\end{figure}

In addition, the present value of the DE EoS is estimated as $w_0 = -0.9753$, which clearly lies in the quintessence regime ($w > -1$) for both the original and reconstructed CC+PPS+DESI DR2 datasets. This indicates a non-cosmological constant behavior of DE, which is slowly evolving in the derived model in the light of data combination used. 
The deviation of $w_{0}$ from $-1$ supports the inference that the model allows a mildly time-dependent DE component capable of driving late-time cosmic acceleration. The ANN-based reconstruction further validates that this behavior remains consistent even when observational uncertainties are accounted for through bootstrap resampling.
One can see the evolution of $w(z)$ with redshift in Figure~\ref{w(z)} along with the $1\sigma$, $2\sigma$, and $3\sigma$ confidence regions. The evolution obtained using the original data is represented in red color, while that derived from the reconstructed data is shown in blue color. The solid curves denote the mean values, and the shaded bands correspond to the respective confidence levels. The curves remain above $w=-1$ throughout the observable range, confirming the quintessence-like behavior of dark energy in our investigation.

The additional freedom introduced by allowing curvature and dynamical DE evolution leads to correlated parameter adjustments when fitting the observational data. In this context, the present value $w_0=-0.9753$ participates in this interplay by slightly modifying the late-time expansion rate, which in turn requires compensating shifts in the matter sector to maintain consistency with observations. Such coupled behavior is expected to extend to structure-formation interpretation, as variations in the expansion history influence the growth of matter perturbations and the inferred clustering amplitude. Nevertheless, because $w_0$ remains close to $-1$, its role is not dominant, and the principal source of degeneracy arises from the combined geometric and dynamical freedom associated with $\Omega_{k0}$ and $\alpha$.

The curvature density parameter is constrained as $\Omega_{k0} = 0.068 \pm 0.029$ from CC+PPS+DESI DR2 (original), indicating a slight preference toward an open Universe at 68\% CL, whereas at 95\% and 99\% CL, $\Omega_{k0}$ is consistent with zero and indicates a flat geometry of the Universe. The ANN-based reconstructed CC+PPS+DESI DR2 dataset yields a tighter constraint with $\Omega_{k0} = -0.131 \pm 0.032$ at 68\% CL, which is statistically sufficient to state a closed geometry of the universe in our derived model. The one-dimensional marginalized distributions for $\Omega_{k0}$ with individual and combined datasets are shown in Figure~\ref{1-dim}.

\begin{figure}[ht]
	\centering
	\includegraphics[scale=0.42]{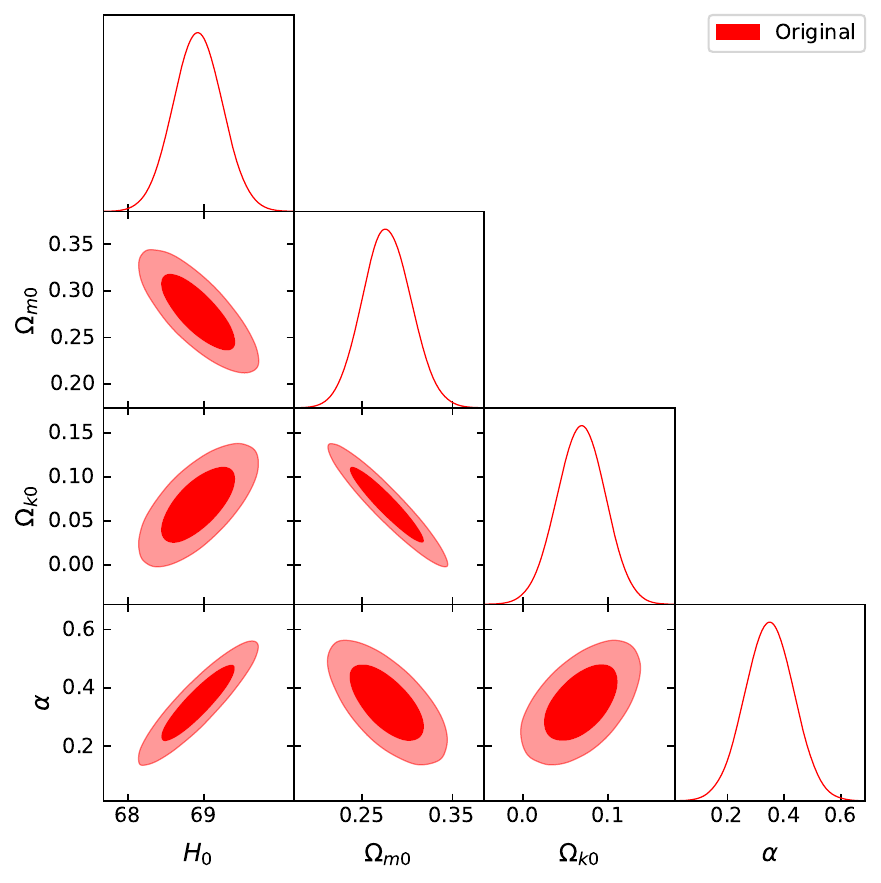}
 	\includegraphics[scale=0.42]{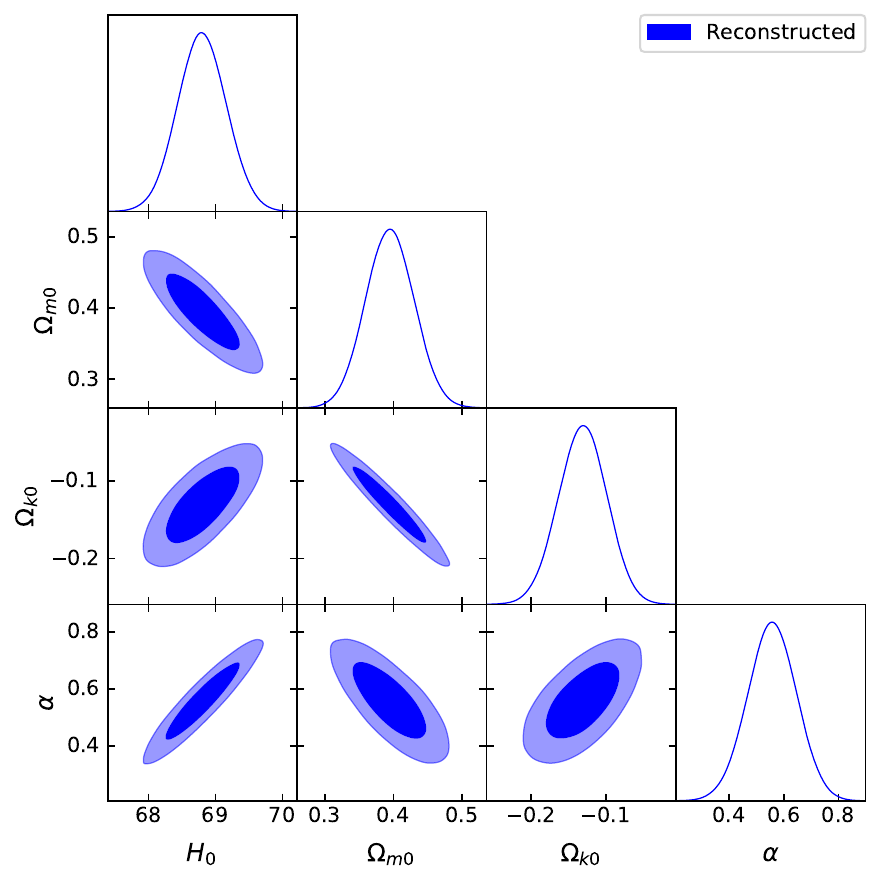}
  	\caption{The likelihood contours at $ 68\% $ and $ 95\% $ confidence levels using original (left panel) and reconstructed (right panel) datasets for our model with: CC+PPS+DESI DR2.} 
	\label{model1}
\end{figure}

The constraints from joint analysis on the Hubble constant $H_0$ show a consistent behavior for both the original and ANN-based reconstructed data in our model. For the original dataset, we obtain $H_0 = 68.92 \pm 0.32$ at 68\% CL, which is in good agreement with Planck CMB measurement ($H_0 = 67.37 \pm 0.54$ $\mathrm{Km\,s^{-1} Mpc^{-1}}$) at 68\% CL within the $\Lambda$CDM framework~\cite{aghanim2020planck}. From the reconstructed dataset, we find almost similar constraints, $H_0 = 68.80 \pm 0.36$ $\mathrm{Km\,s^{-1} Mpc^{-1}}$ at 68\% CL. This indicates that our reconstruction method preserves the statistical behavior of the original data without introducing significant bias in the estimation of the Hubble constant, see Figure~\ref{model1} and Table \ref{tabparm1}. 
The present matter density parameter is $\Omega_{m0} = 0.277 \pm 0.027$, which is close to the estimates from the Planck CMB measurement~\cite{aghanim2020planck}. However, the reconstructed dataset yields $\Omega_{m0} = 0.395 \pm 0.035$, which is relatively higher when compared to the original analysis.  This higher value on $\Omega_{m0}$ from reconstruction indicates a relatively matter-dominated present Universe compared to the standard $\Lambda$CDM expectations. From the total budget relation $(\Omega_{m0}+\Omega_{k0}+\Omega_{\rm DE~0}=1)$, this higher matter fraction implies a reduced contribution from dark energy and/or curvature. This behavior reflects parameter degeneracies among the matter density, curvature, and expansion dynamics, particularly when curvature freedom is introduced and the expansion history is smoothed.
The likelihood contour plots from the joint analysis for the original dataset (left panel) and reconstructed dataset (right panel) are displayed in Figure~\ref{model1}.

To further assess the reliability and interpretive robustness of our framework, we compare our results with those obtained for both the standard $\Lambda$CDM and the $\Lambda$CDM+$\Omega_k$ scenarios using the same datasets as baseline references.
For the flat $\Lambda$CDM model, the constraints on the Hubble constant are significantly higher than those obtained in our parametrized framework, with  $H_0 = 73.40\pm 0.15$ $\mathrm{Km\,s^{-1} Mpc^{-1}}$ (original) and $H_0 = 72.94 \pm 0.15$ $\mathrm{Km\,s^{-1} Mpc^{-1}}$ (reconstructed) at 68\% CL. These values are consistent with the local distance-ladder measurement reported by the SH0ES team, $H_0 = 73.04 \pm 1.04$ $\mathrm{Km\,s^{-1} Mpc^{-1}}$ at 68\% CL~\cite{riess2004type}. Such higher values are expected since spatial flatness is imposed, forcing the expansion history to adjust primarily through the Hubble parameter. In contrast, the additional degree of freedom introduced through the parameter $\alpha$ in our model modifies the DE dynamics and redistributes how the expansion history is accommodated, allowing part of the contribution to be absorbed by the evolving equation of state and thereby yielding comparatively lower $H_0$ estimates. 

\begin{figure}[ht]
	\centering
	\includegraphics[scale=0.42]{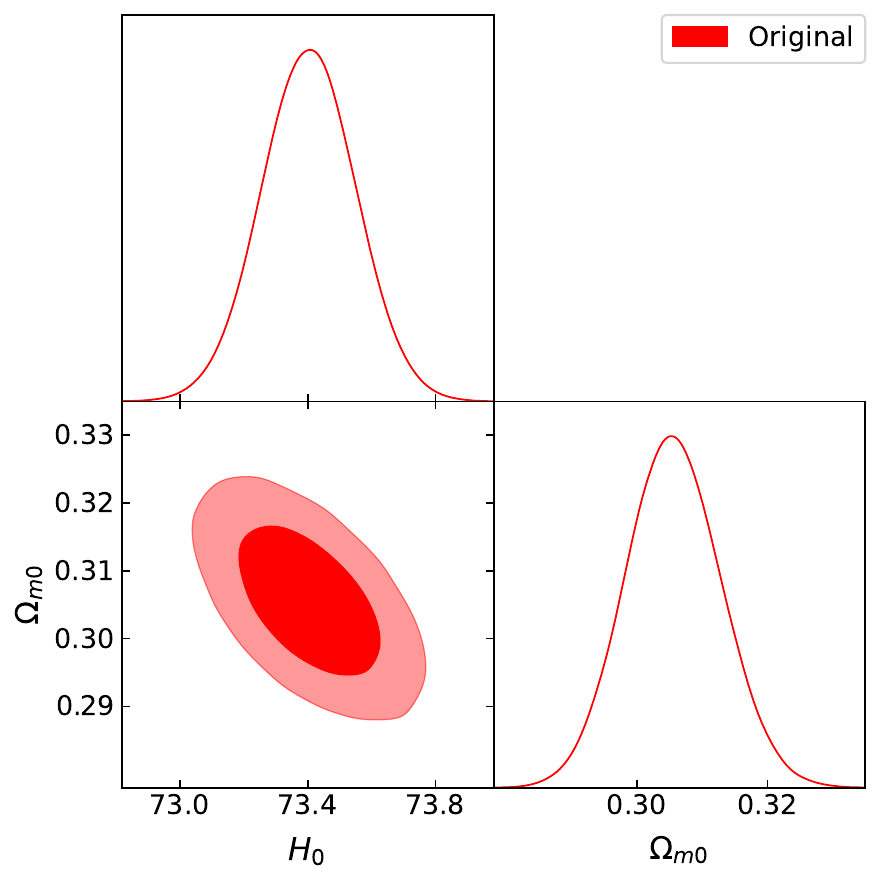}
 	\includegraphics[scale=0.42]{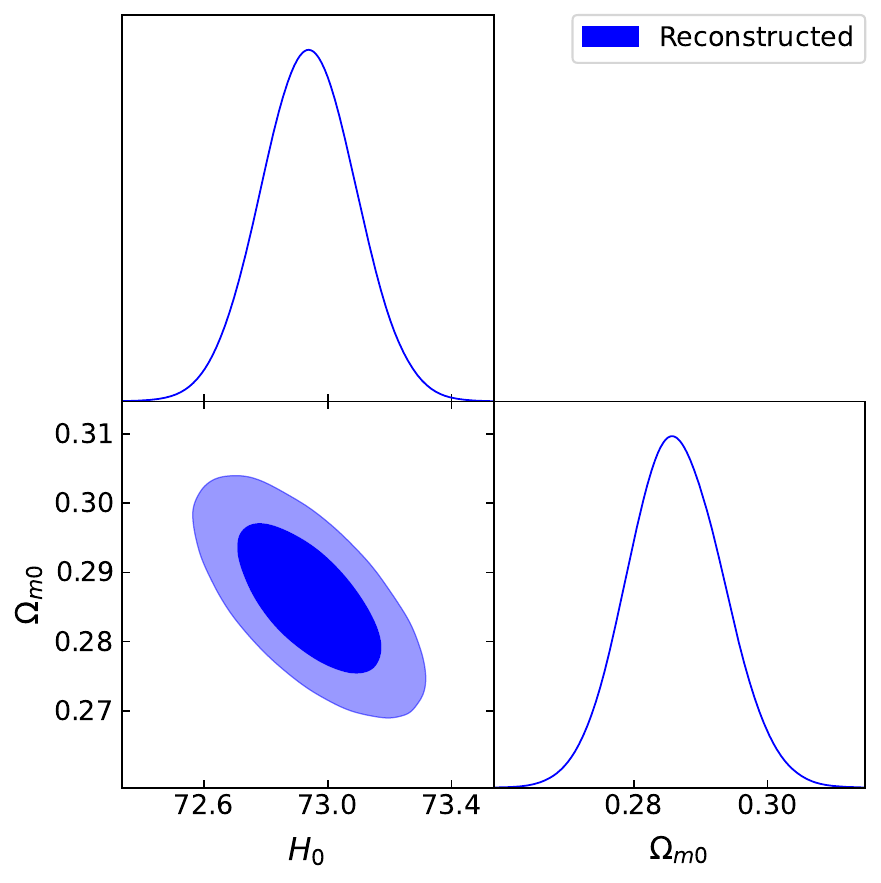}
  	\caption{The likelihood contours, with $ 68\% $ and $ 95\% $ confidence levels using original (left panel) and reconstructed (right panel) datasets for the $\Lambda$CDM model with: CC+PPS+DESI DR2.} 
	\label{model2}
\end{figure}

For the matter density parameter, the $\Lambda$CDM model yields tighter constraints than our parametrized framework, with  $0.3056\pm 0.0073$ (original) and $\Omega_{m0} = 0.286 \pm 0.0072$ (reconstructed). These values remain mutually consistent and lie within the expected range for the standard cosmological scenario. Compared with our results, the original dataset shows good agreement, whereas the reconstructed dataset in our model produces a slightly higher estimate, reflecting the greater parameter freedom that redistributes the relative contributions of matter and DE dynamics. The corresponding likelihood contour plots are shown in Figure~\ref{model2}.

Allowing spatial curvature in the $\Lambda$CDM+$\Omega_k$ scenario leads to a different redistribution of parameter sensitivity. In this case, the Hubble constant constraints remain higher than those obtained in our framework, with $H_0 = 73.33 \pm 0.16$ $\mathrm{Km\,s^{-1} Mpc^{-1}}$ (original) and $H_0 = 72.51 \pm 0.16$ $\mathrm{Km\,s^{-1} Mpc^{-1}}$ (reconstructed) at 68\% CL. This indicates that curvature freedom alone does not substantially lower the inferred expansion rate. Instead, the comparatively lower $H_0$ values in our model arise from the coupled influence of curvature and evolving DE dynamics governed by parameter $\alpha$.
For the matter density parameter, the $\Lambda$CDM+$\Omega_k$ analysis yields $\Omega_{m0} = 0.331 \pm 0.020$ (original) and $\Omega_{m0} = 0.546 \pm 0.024$ (reconstructed). While the original estimate lies within standard cosmological bounds and remains consistent with~\cite{Patil:2024mno}, the reconstructed value shows a noticeable increase. This increase should not be interpreted as unphysical; rather, it reflects parameter degeneracies among matter density, curvature, and expansion dynamics when curvature freedom is introduced and the expansion history is smoothed. In our parametrized model, part of this redistribution is absorbed by the evolving DE sector, resulting in comparatively moderate shifts in $\Omega_{m0}$.

\begin{figure}[ht]
	\centering
	\includegraphics[scale=0.42]{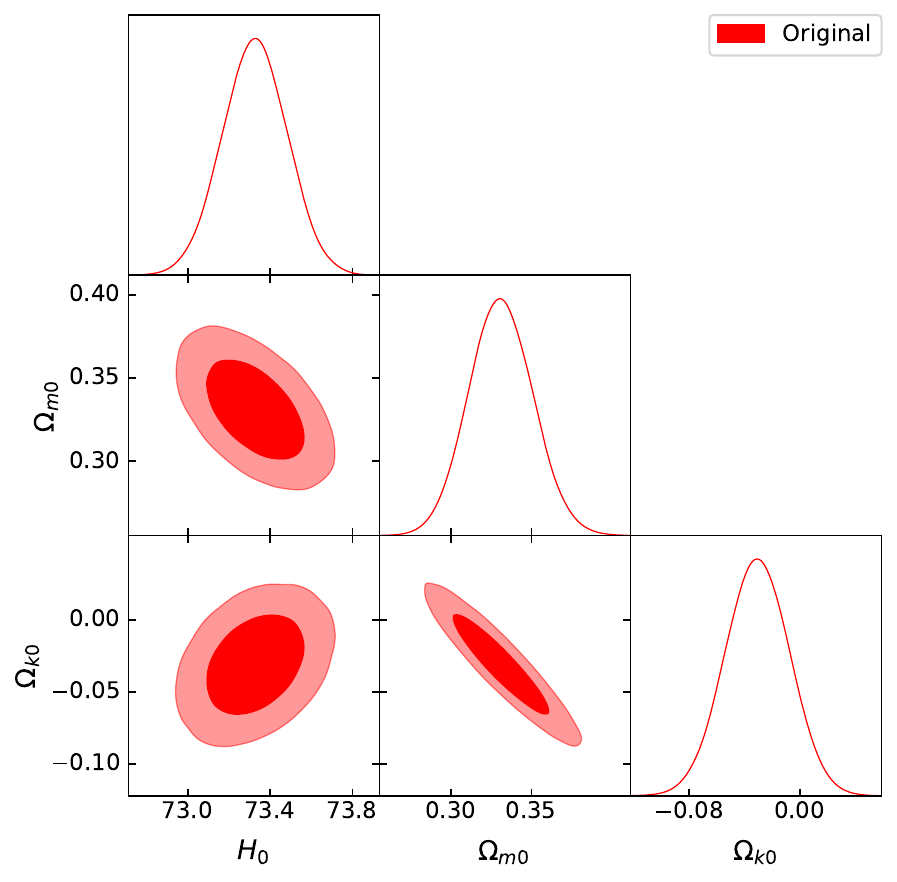}
 	\includegraphics[scale=0.42]{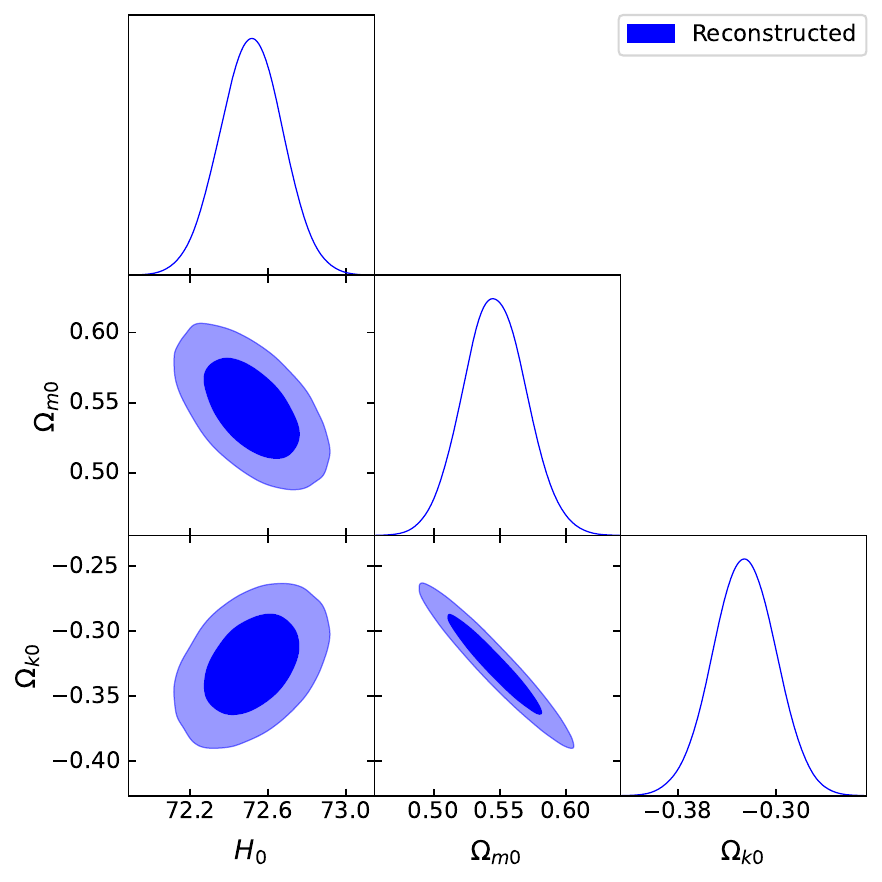}
  	\caption{ The likelihood contours, with $ 68\% $ and $ 95\% $ confidence levels using original (left panel) and reconstructed (right panel) datasets for the $\Lambda$CDM + $\Omega_k$ model with: CC+PPS+DESI DR2.}
	\label{model3}
\end{figure}

The curvature density parameter is constrained as $\Omega_{k0} = -0.031 \pm 0.023$ for the original dataset, which shows a mild preference for closed curvature at 68\% CL, and consistent with a flat Universe within 95\% and 99\% CL. In contrast, the reconstructed dataset favors $\Omega_{k0} = -0.326 \pm 0.026$, indicating a closed geometry. Relative to our model, curvature variations appear more pronounced in $\Lambda$CDM+$\Omega_k$ because geometric freedom serves as the primary mechanism for fitting deviations in the expansion history. In contrast, the presence of $\alpha$ distributes this adjustment between curvature and DE evolution, leading to a distinct degeneracy structure among cosmological parameters.
The likelihood contour plots for the original and reconstructed datasets for the $\Lambda$CDM+$\Omega_k$ model are displayed in Figure~\ref{model3}. Overall, comparison with both baseline scenarios demonstrates that the parameter shifts observed in our framework are not purely geometric artifacts but are influenced by the dynamical DE component, reinforcing the robustness and physical interpretability of the proposed model.

The ANN-based reconstruction provides physical interpretive value in the present study. By generating a smooth representation of the expansion history that propagates observational uncertainties through bootstrap resampling, it allows us to evaluate the sensitivity of curvature and DE constraints to fluctuations in sparsely sampled data. In particular, comparing results from original and reconstructed datasets enables us to identify re-distribution of parameter sensitivity among curvature, matter density, and the DE evolution parameter, thereby offering insight into degeneracy structure that cannot be inferred from a single analysis alone. It is important to mention here that the constraints obtained from the ANN reconstruction do not present significant evidence that it improves the agreement between competing cosmological frameworks.

\subsection{Model Comparision}
 Further, we performed a chi-square $\chi^2$ analysis to test the goodness of fit for all three models with the original and reconstructed datasets. Also, the reduced chi-square $\chi^2_r$ values are calculated to test the model consistency. We calculated the minimum chi-square $\chi^2_{\rm min}$  by using the relation $\chi^2_{\rm min}=-2 \times \log \textrm{(Max Likelihood)}$. 
The reduced chi-square $\chi^2_r$ can be obtained by dividing $\chi^2_{\rm min}$ by the degrees of freedom, given by observational data points used in our analysis minus the number of model parameters.  In the present analysis, we used $N=1741$ data points for both the original and reconstructed CC+PPS+DESI DR2 datasets. The $\chi^2_{r}$ values can be interpreted as: if $\chi^2_{r} \approx 1$, it indicates the model fits well; if it deviates significantly from $\chi^2_{r} \approx 1$ indicate underfitting $(\chi^2_{r} \gg 1)$ or overfitting $(\chi^2_{r} \ll 1)$ as the case may be. The values of $\chi^2_{r}$ in our analysis are displayed in Table~\ref{tabparam2}. 

\begin{table}[ht]
\caption{Side-by-side model comparison statistics using the CC+PPS+DESI DR2 dataset.}
\label{tabparam2}
\renewcommand{\arraystretch}{1.6}
\begin{tabular}{l l c c c}
\toprule
\textbf{Data} & \textbf{Statistic} & \textbf{$\Lambda$CDM} & \textbf{$\Lambda$CDM+$\Omega_k$} & \textbf{Our model} \\
\hline \hline

\multirow{4}{*}{Original}
&  \boldmath{$\chi^2_{\min}$} & 1801 & 1800 & 1792 \\
&  \boldmath{$\chi^2_{\rm r}$} & 1.04 & 1.04 & 1.03 \\
& \textbf{AIC}              & 1807 & 1807 & 1802 \\
& \textbf{BIC}              & 1823 & 1829 & 1830 \\

\midrule

\multirow{4}{*}{Reconstructed}
&  \boldmath{$\chi^2_{\min}$} & 2068 & 1884 & 1829 \\
&  \boldmath{$\chi^2_{\rm r}$} & 1.19 & 1.08 & 1.05 \\
& \textbf{AIC}              & 2074 & 1892 & 1839 \\
& \textbf{BIC}              & 2090 & 1914 & 1867 \\

\botrule
\end{tabular}
\end{table}

We also evaluated the model performance using the Akaike Information Criterion (AIC)~\cite{akaike2003new,burnham2011aic} and Bayesian Information Criterion (BIC)~\cite{schwarz1978estimating}. These criteria provide a robust statistical framework for model comparison by penalizing the number of free parameters to avoid overfitting. The AIC is calculated as
$\mathrm{AIC} = \chi^2_{\text{min}} + 2k$,
where $k$ is the number of model parameters. Similarly, the BIC is given  $\mathrm{BIC} = \chi^2_{\text{min}} + k\ln(N)$,
where $N$ is the number of observational data points. In both cases, lower AIC and BIC values indicate a statistically preferred model. The corresponding AIC and BIC values obtained from our analysis are summarized in Table~\ref{tabparam2}.

From the table, it is observed that our model provides lower values of $\chi^2_{\text{min}}$, and AIC compared to the standard $\Lambda$CDM scenario for the original dataset, indicating a comparatively better goodness of fit.  When compared with the $\Lambda$CDM+$\Omega_k$ scenario, our model also yields lower $\chi^2_{\text{min}}$ and AIC values, favoring our model due to the improved fit achieved without excessive parameter penalization. The reduced chi-square values remain very close to unity for all models, confirming the statistical consistency of the fits. For the reconstructed dataset, although the $\chi^2_{\text{min}}$, AIC, and BIC values increase slightly due to reconstruction propagation, our model continues to yield relatively lower values than $\Lambda$CDM  and $\Lambda$CDM+$\Omega_k$. This result suggests that even after data reconstruction, our framework maintains a competitive statistical performance.

\section{Conclusion}\label{section5}
In this work, we have derived observational constraints on DE EoS parametrization in a non-flat cosmological framework. We have employed recent low-redshift datasets, including CC, PPS, and DESI DR2, and constrained the model parameters with original data and ANN-based reconstructed data. The main numerical outcomes of our analysis using the CC+PPS+DESI DR2 dataset are summarized as follows:
\begin{itemize}

\item For our DE model with curvature, the original dataset yields
$H_0 = 68.92 \pm 0.32$,
$\Omega_{m0} = 0.277 \pm 0.027$,
$\Omega_{k0} = 0.068 \pm 0.029$,
and $\alpha = 0.348 \pm 0.087$ (68\% CL), indicating a mildly open geometry. Using the reconstructed dataset, the same model gives
$H_0 = 68.80 \pm 0.36$,
$\Omega_{m0} = 0.395 \pm 0.035$,
$\Omega_{k0} = -0.131 \pm 0.032$,
and $\alpha = 0.557 \pm 0.089$ (68\% CL), corresponding to a closed spatial geometry and an increased matter density.

\item We have noticed that the transition from original to reconstructed data shifts $\Omega_{m0}$ upward by $\sim 43\%$ and reverses the sign of $\Omega_{k0}$, while simultaneously increasing $\alpha$, suggesting a correlated response between DE evolution and spatial curvature. The EoS parameter value is estimated as $ w_0 =- 0.9753$, indicating a quintessence-like behavior of DE for both the original and reconstructed datasets. 

\item In the flat $\Lambda$CDM scenario, the original analysis yields
$H_0 = 73.40 \pm 0.15$ and
$\Omega_{m0} = 0.3056 \pm 0.0073$,
while the reconstructed dataset gives
$H_0 = 72.94 \pm 0.15$ and
$\Omega_{m0} = 0.286 \pm 0.0072$,
showing comparatively smaller shifts relative to our model. The non-flat $\Lambda$CDM scenario produces
$\Omega_{k0} = -0.031 \pm 0.023$ (original) and
$\Omega_{k0} = -0.326 \pm 0.026$ (reconstructed),
with corresponding increases in $\Omega_{m0}$ from $0.331 \pm 0.020$ to $0.546 \pm 0.024$.

\item Across all models, reconstructed datasets consistently favor higher matter density and more negative curvature, while $H_0$ remains comparatively stable within each framework but differs systematically between our DE model ($H_0 \simeq 68.8$ km\,s$^{-1}$\,Mpc$^{-1}$) and $\Lambda$CDM ($H_0 \simeq 73$ km\,s$^{-1}$\,Mpc$^{-1}$).

\item These quantitative trends indicate that the reconstructed data enhance correlations between $\Omega_{m0}$, $\Omega_{k0}$, and the DE evolution parameter $\alpha$, highlighting the sensitivity of curvature constraints to both the DE dynamics and the data reconstruction procedure.

\item The statistical comparison indicated that the proposed parametrization provides a competitive fit to current low-redshift datasets when analyzed within a non-flat spacetime framework. The ANN-based reconstruction can be primarily viewed as a complementary consistency check on parameter inference under smoothed data realizations.
\end{itemize}

Future cosmological surveys such as DESI full data releases~\cite{DESI:2025fxa, Spec-S5:2025uom}, Euclid~\cite{Euclid:2024yrr}, and Large Synoptic Survey Telescope(LSST)~\cite{LSST:2008ijt},  will significantly improve measurements of the expansion history and large-scale structure. At the same time, CMB-S4 observations~\cite{CMB-S4:2016ple} will greatly improve constraints on early-universe physics and help reduce parameter degeneracies when combined with late-time cosmological data. These datasets may allow stronger tests of curvature–DE degeneracies and provide opportunities to further assess phenomenological parametrizations such as the one explored here.

\bigskip
\noindent\textbf{Acknowledgments}
We are grateful to the honorable referee and to the editor for the constructive comments/suggestions that have significantly improved the work in terms of quality and presentation.

\bigskip
\noindent\textbf{Author contributions}
DRK:. Writing - original draft preparation; Methodology. SKY:. Formal analysis and investigation; Writing - review and editing; Supervision. Both authors reviewed and approved the final manuscript.

\bigskip
\noindent\textbf{Funding information} The authors did not receive support from any organization for the submitted work.

\bigskip
\noindent\textbf{Data availability statement} No datasets were generated or analysed during the current study.

\bigskip
\noindent\textbf{Conflicts of interest} The authors declare no conflict of interest.\\

\section*{Declarations}
\noindent\textbf{Ethics declaration} Not applicable.

\bigskip
\noindent\textbf{Competing interests} The authors declare no competing interests.

\bibliography{refs}
\end{document}